\documentclass{aa}  
\usepackage{graphicx}
\usepackage{txfonts}
\begin{document}
   \title{Planetary companions around the metal-poor star \mbox{HIP 11952}}


   \author{J. Setiawan\inst{1}
          \and
          V. Roccatagliata\inst{2,1}
          \and
          D. Fedele\inst{3}
          \and
          Th. Henning\inst{1}
          \and
          A. Pasquali\inst{4}
          \and
          M.V. Rodr{\'i}guez-Ledesma\inst{1,5}
	  \and
          E. Caffau\inst{4}
	  \and
          U. Seemann \inst{5,6}
	  \and
          R.J. Klement \inst{7,1}
           }

   \offprints{J. Setiawan}

   \institute{Max-Planck-Institut f\"ur Astronomie, K\"onigstuhl 17, D-69117
   Heidelberg, Germany\\
              \email{setiawan@mpia.de}
         \and
             Space Telescope Science Institute, 3700 San Martin Drive, Baltimore, MD 21218, USA\\
         \and
	     Department of Physics and Astronomy, Johns Hopkins University, 3400 North Charles Street, Baltimore, MD 21218, USA\\
         \and
             Astronomisches Rechen-Institut, Zentrum f\"ur Astronomie, M\"onchhofstrasse, 12-14, D-69120, Heidelberg\\
         \and
	     Institut f\"ur Astrophysik, Georg-August-Universit\"at, Friedrich-Hund-Platz 1, D-37077 G\"ottingen, Germany\\
         \and
             European Southern Observatory, Karl-Schwarzschild Str. 2, D-85748, Garching bei M\"unchen, Germany\\
         \and
            University of W{\"u}rzburg, Department of Radiation Oncology, D-97080 W{\"u}rzburg, Germany\\
	    }

   \date{Received 4/8/2011 - Accepted 27/2/2012}

 
  \abstract
   {}
   {We carried out a radial-velocity survey to search for planets around metal-poor stars. 
   In this paper we report the discovery of two planets around \mbox{HIP 11952}, a metal-poor star with [Fe/H]$=-1.9$ 
   that belongs to our target sample.}
  {Radial velocity variations of \mbox{HIP 11952} were monitored systematically 
   with FEROS at the 2.2~m telescope located at the ESO La Silla observatory from August 
   2009 until January 2011. We used a cross-correlation technique to measure the stellar radial velocities (RV).}
  {We detected a long-period RV variation of 290 d and a short-period one of 6.95 d. 
  The spectroscopic analysis of the stellar activity reveals a stellar rotation period of 4.8 d. 
  The Hipparcos photometry data shows intra-day variabilities, which give evidence for stellar pulsations. 
  Based on our analysis, the observed RV variations are most likely caused by the presence 
  of unseen planetary companions. Assuming a primary mass of 0.83 M$_\odot$, we computed minimum planetary masses 
  of 0.78 M$_\mathrm{Jup}$ for the inner and 2.93 M$_\mathrm{Jup}$ for the outer planet. The semi-major axes are 
  $a_1=0.07$ AU and $a_2=0.81$ AU, respectively.}
  {\mbox{HIP 11952} is one of very few stars with [Fe/H]$<-1.0$ which have
   planetary companions. This discovery is important to understand planet formation around metal-poor stars.}

   \maketitle
%

\section{Introduction}

Current results of the exoplanet surveys strongly suggest a correlation 
between a star's stellar metallicity and its probability of hosting planets, 
in particular for main-sequence stars (e.g., \cite{fischer05} 2005; \cite{johnson10}).
According to  these studies, the detection rate of planets decreases 
with metallicity. However, the conclusions of \cite{fischer05} (2005) might be affected by an
observational bias, since these authors did not have similar numbers of stars 
in their survey per bin of metallicity. 
Therefore, it is crucial to understand if either the high
stellar metallicity triggers planet formation or the metal enhancement
of stars is caused by the formation of planets.   

In the last years, exoplanet surveys tried to bridge this gap, starting to include more
metal-poor stars in their samples. \cite{sozzetti09} conducted a three-year RV survey 
to look for planets around metal-poor stars down to [Fe/H]$=-2.0$ and found no evidence for short-period 
giant planets within 2~AU from the central star. \cite{santos11} performed a similar survey, only down to 
[Fe/H]$=-1.4$ and found three moderately metal-poor stars hosting a long period giant planets (P$>$ 700 d). 
A hot Saturn and a hot Jupiter have been found transiting two moderate metal-poor
stars, with [Fe/H]$=-0.46$ and $-0.4$, respectively (\cite{bouchy10}; \cite{simpson11}). 

In June 2009 we started a survey to search for planets around metal-poor stars. The target sample 
includes 96 metal-poor A and F stars. Our target list includes stars with metallicities in the range $-4.0\le$[Fe/H]$\le -0.5$.
As part of this work, \cite{setiawan10} found a planet around an extremely
metal-poor red horizontal branch star with a short period of 16.2 d. 
We notice that its [Fe/H]$=-2.1$ is not included in the metallicity range covered by the surveys of 
\cite{sozzetti09} and \cite{santos11}. 

\begin{figure}[t]
\centering
\includegraphics[height=9.0cm,width=11cm,angle=90]{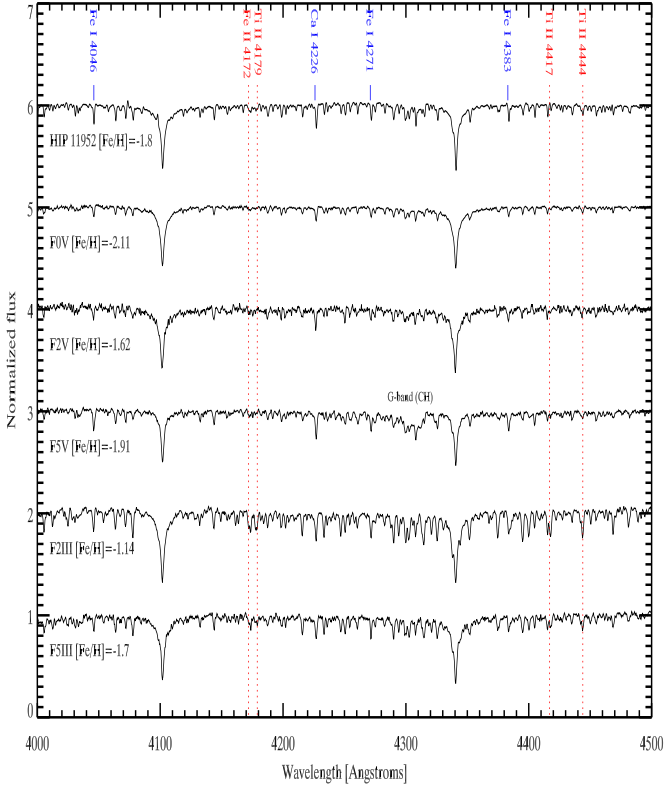}
\caption{Comparison of the spectrum of \mbox{HIP 11952} with stellar spectra of the Indo-US library. 
  The dashed lines highlight some iron lines (\ion{Fe}{ii}, \ion{Ti}{ii}) which are sensitive 
  to the luminosity class of the star. The solid lines in the upper part of 
  the figure indicate the position of some iron and calcium lines which do not vary with the luminosity class.}
\label{spt}
\end{figure}           

These recent observations have started to disclose the realm of planets at rather low 
stellar metallicities, indicating that metallicity may not be the main driver of planet formation. 
Clearly, more statistics is needed to obtain robust conclusions. 
In this framework, we report the detection of two planetary companions around \mbox{HIP 11952} 
as a result of our RV planet survey around metal-poor stars.

The paper is organized as follows: Observations and data reduction are presented in section~\ref{obs}.   
The stellar parameters of \mbox{HIP 11952} are shown in section~\ref{star} together with the most 
relevant information available for this star in the literature. 
In section~\ref{rvobs} we describe the RV and photometric analysis. 
The detection of the planetary companion is addressed in section~\ref{planet}. 
Discussion and conclusions are given in sections \ref{discussion} and 
\ref{con}, respectively.


 \section{Observations and data reduction}  
\label{obs}
We observed \mbox{HIP 11952} from August 2009 until January 2011 with FEROS at the 2.2~m 
Max-Planck Gesellschaft/European Southern Observatory (MPG/ESO) telescope,
located at ESO-La Silla observatory, during the MPG guarantee time. 
FEROS has a resolution of $R=48\,000$ and a wavelength coverage 
of 370--920 nm (\cite{kaufer98}). 
The long-term RV precision of FEROS was measured across a 
period of 6 years, from December 2003 until January 2010, 
using the standard star $\tau$ Ceti (\mbox{HD 10700}). 
We obtained an RV precision of better than 11 $\mathrm{m\,s}^{-1}$.

The data reduction was performed with a software package in the ESO-MIDAS environment available 
online at the telescope. 
The procedure of the RV computation from the FEROS spectra is described in 
\cite{setiawan03} and based on a cross-correlation technique. 
The spectra of \mbox{HIP 11952} were cross-correlated with a template of an F dwarf star 
which best matches the spectral type of \mbox{HIP 11952}.
      
\section{Analysis}
\label{an}
In this section we present the analysis of the stellar parameters 
and RVs derived from the FEROS spectra. 

\subsection{The star \mbox{HIP 11952}}
\label{star}   
 
\mbox{HIP 11952} (\mbox{HD 16031}; \mbox{LP 710-89}) was previously classified as an F0 dwarf star 
(e.g., \cite{wright03}; \cite{Kharchenko09}) and as a giant G8 (e.g. \cite{Sanchez-Blazquez06}; \cite{Cenarro07}). 

Our spectral classification was carried out by comparing the FEROS spectrum of \mbox{HIP 11952} 
with spectra from the Indo-US library (\cite{Valdes2004}) of metal-poor stars with different 
luminosity classes and spectral types.  
The FEROS spectrum was convolved to the resolution of the Indo-US spectral library
(1\AA). 
Following the spectral classification criteria presented by \cite{GrayCorbally2009}, 
we used  some lines  (\ion{Fe}{ii}, \ion{Ti}{ii}) which are sensitive to the luminosity class of the star 
as well as other iron (\ion{Fe}{i}) and calcium (\ion{Ca}{i}) lines 
which do not vary with the luminosity class (Fig.~\ref{spt}). 
From this comparison we concluded that \mbox{HIP 11952} is an F2V star. 
However, this result has to be confirmed by an independent 
spectroscopic analysis of the stellar spectra, as we present below.
            
\mbox{HIP 11952} is at a distance of 115.3 pc as derived from the 
parallax measurements given in the Hipparcos catalogue (\cite{perryman97}).  
Fundamental parameters of this star were determined using our high-resolution FEROS spectra. 
In particular, stellar abundances, effective temperature and surface gravity were computed using 
the synthetic spectra from the 1D ATLAS models (\cite{kurucz93}; \cite{kurucz05}) and the fit of the H$\alpha$ line 
to the CO5BOLD 3D model atmosphere (\cite{caffau2011}).    
The atomic data of the iron lines are from
the Large Program "First Stars" lead by R. Cayrel, optimized
for metal-poor stars (see \cite{sivarani04}). 
The first attempt of abundance analysis was based on 1D ATLAS model 
atmospheres computed using the version 9 of the code ATLAS 
(\cite{kurucz93}; \cite{kurucz05}) running under Linux (\cite{sbordone04};
\cite{sbordone05}).
We derived a temperature of 5960\,K by fitting
the H$\alpha$ wings with a grid of synthetic spectra computed from
ATLAS models, and a temperature of 6120~K when we used a
grid of synthetic spectra based on 3D models.

By imposing an agreement between the iron
abundance derived from the lines of \ion{Fe}{i} and the lines of \ion{Fe}{ii}, 
we derived surface gravities $\log\,g$ of 3.8 and 4.0 for the two cases 
of T$\_\mathrm{eff}=5960$\,K and 6120\,K, respectively. The uncertainty on the 
effective temperature derived from the fit of the H$\alpha$ is 150\,K, 
while the error in the surface gravity is 0.3.
A microturbulence of 1.4~$\mathrm{km\,s}^{-1}$ was derived by minimizing the slope of the
abundance versus equivalent width (EW) relation.
The resulting iron abundance derived is [Fe/H]$=-1.95\pm0.09$ for ${\rm T_\mathrm{eff}=5960}$\,K 
and [Fe/H]$=-1.85\pm0.09$ for ${\rm T_\mathrm{eff}=6120}$\,K, respectively.

We thus obtained two possible parameter sets for this star:
(${\rm T_{\rm eff}}$, $\log g$, [Fe/H])=(5960~K, 3.8,$-1.95$) (1D ATLAS) and (6120~K, 4.0, $-1.85$) (3D MODELS).
We measured an EW=3.07$\pm$0.03~m\AA~of the Li feature at 670.7~nm,
which implies an abundance A(Li)=2.2 if we fix ${\rm T_{\rm eff}=6120}$\,K,
and A(Li)=2.1 in the case ${\rm T_{\rm eff}=5960}$\,K.
       
\cite{feltzing2001} derived a stellar age of $12.8\pm2.6$ Gyr by comparing the
Str\"omgren photometry of \mbox{HIP 11952} with the evolutionary tracks computed 
for the Str\"omgren metallicity ([m/H]= -1.54) of the star. 
We checked their result using the photometry provided by Hipparcos for
\mbox{HIP 11952} in the Bessell filters system (\cite{bessel2000}) with the isochrones by 
\cite{marigo2008} and \cite{girardi2010}, calculated for the Str\"omgren metallicity of the star. 
In the assumption that the dust extinction along the line of sight to \mbox{HIP 11952} is not large, the
Hipparcos photometry indicates, within its uncertainty, an age older than 10 Gyr for this star.

We used the isochrones by \cite{marigo2008} and \cite{girardi2010} computed for [m/H]= -1.54 
and ages in the range 10--13 Gyr in order to consistenly derive the mass and radius of \mbox{HIP 11952}. 
By comparing its photometry with the selected isochrones, we constrained its mass
between 0.79~M$_\odot$ and 0.88~M$_\odot$.
From the relation between surface gravity, stellar mass and stellar radius:
\begin{equation}
\frac{R}{R_\odot}=\sqrt{\frac{M}{M_\odot}\frac{g_\odot}{g}}
\end{equation}
we derived a stellar radius of 1.6$\pm$0.1~R$_\odot$, where the error is obtained from
the errors propagation. The derived stellar mass and radius adopted in this work are: 
$0.83^{+0.05}_{-0.04}$~M$_\odot$ and 1.6$\pm$0.1~R$_\odot$.

Fundamental stellar parameters have also been estimated in previous studies (e.g., \cite{masana06}; \cite{charbonnel05}). 
The corresponding values are ${\rm T_{\rm eff}=6367}$~K, a stellar radius $R_*=1.0$~R$_\odot$ and 
surface gravity $\log g=4.1$. \cite{feltzing2001} reported $m=0.785$~M$_\odot$. 
If we compile all literature values for the surface gravity
given in \cite{cayrel01} and use $R_*=1.0$~R$_\odot$, we obtain a mean value $m=0.55\pm0.23$~M$_\odot$. 
Within the errors, these values are consistent with those we derived, although the stellar radius is smaller than the
one we adopted. Finally, we compared the derived stellar parameters with those given by \cite{casagrande2010}. 
These authors measured $\log\,g=$ 4.17, [Fe/H]=$ -1.74$ and T$_\mathrm{eff}=$ 6186~K.
Within the uncertainties, these values are in good agreement with our determination.

Based on the surface gravity and radius derived from our 
analysis and the available literature data, \mbox{HIP 11952} is more likely a star
already evolved off the main-sequence, roughly sitting at the base of the 
subgiant branch.
 
Astrometric and photometric data of \mbox{HIP 11952} can be found 
in public catalogues (e.g., Hipparcos).  
The astrometric variability is less than 5 mas, allowing the conclusion 
that an unseen stellar companion in the system can be ruled out. 
Nevertheless, older RV measurements from \cite{eggen59} combined with \cite{carney87}  
suggested that \mbox{HIP 11952} may be a spectroscopic binary SB1 (\cite{fouts87}). 
However, our measurements can neither reject nor confirm this claim yet.  
The updated parameters of \mbox{HIP 11952} are given in Table~\ref{par}.

\mbox{HIP 11952} is listed as a member of the metal-poor stellar stream 
detected by \cite{arifyanto06}, a group of
putative thick disk stars moving on similar orbits in the Galactic potential and 
currently lagging the Local Standard of Rest by $V_\mathrm{lag}\approx80$ km s$^{-1}$.
As such, \mbox{HIP 11952} might stem from one of the Milky 
Way's former satellite galaxies that once got tidally disrupted. 
Alternatively, \cite{minchev09} showed that 
the stream might consist of \textit{in situ} disk stars being scattered 
towards common orbits through a merger-induced perturbation of the old 
disk $\sim1.9$ Gyr ago. It is also possible that there is a connection to 
the Arcturus stream at $V_\mathrm{lag}\approx100$ which itself could 
have a tidal or a dynamical origin (see
discussion in \cite{klement10} for more details).

\begin{figure}[t]
\centering
\includegraphics[width=8.5cm]{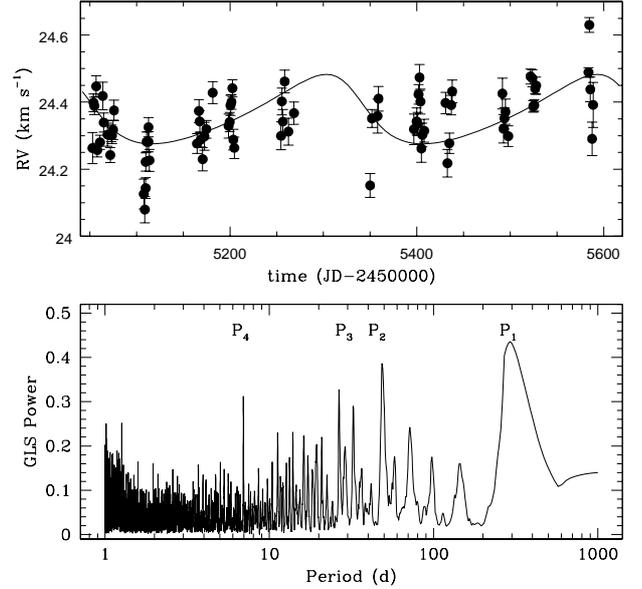}
\caption{RV observations of \mbox{HIP 11952}, taken with FEROS 
from August 2009 until January 2011 (upper panel). 
We calculated a single Keplerian fit to the data (solid line) for the outer component.
A multi-component Keplerian fit is presented in Fig.\ref{orbkepler}, see Table.~\ref{orbsol}. 
The lower panel shows a Generalized Lomb-Scargle (GLS) periodogram of the whole RV data set. 
The highest peak, marked with P$_1$ corresponds to a period of $290\pm16$ d with 
a False Alarm Probability of $\sim 7\times10^{-6}$.}
\label{rv}
\end{figure}

\begin{table}[h]
\caption{Stellar parameters of \mbox{HIP 11952}} 
\label{par}
\vspace{-0.5cm}
$$
\begin{array}{lll}
\hline
\mathrm{Parameter}	& \mathrm{value}  		& \mathrm{unit} \\
\hline

\mathrm{Spectral\,\,type}^{*}		& \mathrm{F2V-IV}      		&  \\
m_{V}   	^{(a)} 			& \mathrm{9.88}\pm0.02  	& \mathrm{mag} \\
\mathrm{Parallax}^{(a)} 		& \mathrm{8.67}\pm1.81  	& \mathrm{mas} \\
T_{\mathrm{eff}}^{*}			& \mathrm{5960}\pm150  	& K   \\
					& \mathrm{6120}\pm150  	& K   \\
\mathrm{log\,g}	^{*}			& \mathrm{3.8}\pm0.3		& \mathrm{cm}^2/g \\
					& \mathrm{4.0}\pm0.3		& \mathrm{cm}^2/g \\
\mathrm{[Fe/H]}	^{*}			& \mathrm{-1.95}\pm0.09	& \mathrm{dex}  \\
					& \mathrm{-1.85}\pm0.09	& \mathrm{dex}  \\
\mathrm{R}_{*} ^{*}			& \mathrm{1.6}\pm0.1 	& \mathrm{R}_\odot  \\
\mathrm{Mass}^{*}			& \mathrm{0.83}^{+0.05}_{-0.04}	& \mathrm{M}_\odot \\
\mathrm{Age}^{(b)}			& \mathrm{12.8}\pm2.6	& \mathrm{Gyr}  \\
v_{\mathrm{rot}}\sin{i}^{*}		& \mathrm{5.2}\pm1.0	& \mathrm{km\,s}^{-1}  \\
P_{\mathrm{rot}}/\sin{i}  		& \mathrm{15.7}\pm2.5	& \mathrm{d}  \\
\hline
\hline
\end{array}
$$
$^{*}$ this work \\
$^{a}$ Hipparcos catalogue (\cite{perryman97}) \\
$^{b}$ \cite{feltzing2001} \\

\end{table}

\subsection{Radial velocity}
\label{rvobs}
The RV variation of \mbox{HIP 11952} is shown in Fig.~\ref{rv} (upper panel). During the observation
campaigns we obtained 77 RV measurements (Table.~\ref{tabrv}).
We applied the Generalized Lomb-Scargle (GLS) periodogram (\cite{zk09}) to the RV data 
in order to search for periodicities. 

We found several signals in the periodogram, as shown in the lower panel of Fig.~\ref{rv}.
The four highest signals are marked with P$_1$, P$_2$, P$_3$ and P$_4$.    
The highest peak P$_1$ has a False Alarm Probability (FAP) of  $7.2 \times 10^{-6}$ and corresponds 
to a period of $290\pm16$ d. 
We computed a Keplerian fit to the data, shown as the solid line in the upper panel of Fig.~\ref{rv}. 
The parameters of this fit are given in Table~\ref{orbsol}.
 
\begin{figure}[t]
\centering
\includegraphics[width=8.5cm, height=8.5cm]{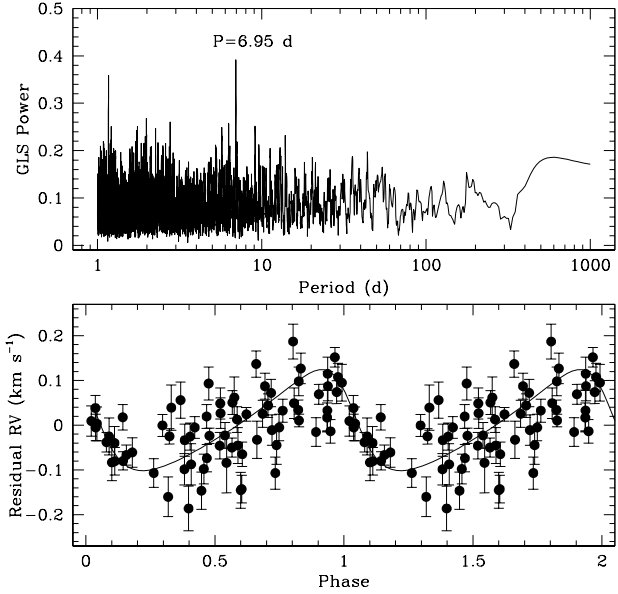}
\caption{The residual RV after removing the 290 d periodicity. The periodogram in the
upper panel shows two peaks at 6.95 d and 1.16 d. The 1.16 d is identified as a harmonic 
of 6.95 d (1/1.16 + 1/6.95 = 1.0). The lower panel shows the phase folded RVs with 
a period of 6.95 d. The solid line shows a Keplerian fit for the residual RV variation.}
\label{resRV}
\end{figure}

After removing the 290 d signal, the peaks P$_2$ and P$_3$ disappear. However, the peak P$_4$ 
remains, which means that this signal is not an alias of $P_1$. 
The signal P$_4$ corresponds to a period of 6.95 d with FAP$=8 \times 10^{-4}$.
In the residual RV periodogram, we also observed a signal at 1.16 d (Fig.~\ref{resRV} upper panel), 
which is obviously a harmonic of the 6.95 d period (1/6.95 + 1/1.16 = 1.0). 
In the lower panel of Fig.~\ref{resRV} we have phase folded the RVs with $P=6.95$ d and show a 
Keplerian fit with the parameters listed in Table~\ref{orbsol}.
After excluding stellar activity as the cause for the RV variations in the next sections, we are going
to interpret the signals $P_1$ and $P_4$ as orbital periods of unseen low-mass companions (section~\ref{planet} below).
	 
\begin{table*}[t]
\caption{RV variation of \mbox{HIP 11952}} 
\label{tabrv}
$$
\begin{array}{lrlllrl}
\hline
\hline
\mathrm{JD}	 & \mathrm{RV}  & \mathrm{error} & \vline & \mathrm{JD}	 &\mathrm{RV}  & \mathrm{error} \\
\mathrm{-2450000}& \mathrm{(m/s)}  & \mathrm{(m/s)} & \vline & \mathrm{-2450000}& \mathrm{(m/s)}  & \mathrm{(m/s)} \\

\hline
5052.8320  &  24262.99 & 46.17 &\vline &  5256.5331  &  24342.01  & 35.92   \\
5053.9449  &  24398.12 & 26.57 &\vline &  5258.5279  &  24461.25  & 33.87   \\
5054.9025  &  24388.69 & 26.20 &\vline &  5262.5458  &  24312.08  & 39.98   \\
5056.7834  &  24447.32 & 31.11 &\vline &  5268.5034  &  24367.11  & 33.39   \\
5057.9442  &  24257.13 & 20.34 &\vline &  5349.9353  &  24151.48  & 35.26   \\
5060.8739  &  24280.16 & 21.22 &\vline &  5351.9288  &  24351.42  & 26.78   \\
5063.8340  &  24418.54 & 41.10 &\vline &  5357.9251  &  24358.36  & 50.76   \\
5064.8816  &  24338.95 & 28.32 &\vline &  5358.9292  &  24410.18  & 36.79   \\
5068.8975  &  24302.92 & 34.02 &\vline &  5396.8522  &  24319.13  & 45.60   \\
5071.9278  &  24242.14 & 22.27 &\vline &  5399.8478  &  24342.35  & 40.48   \\
5072.9043  &  24306.86 & 24.57 &\vline &  5400.9186  &  24334.94  & 34.61   \\
5073.7695  &  24300.97 & 24.45 &\vline &  5401.8807  &  24423.26  & 29.23   \\
5074.9106  &  24317.19 & 28.81 &\vline &  5402.8758  &  24473.63  & 38.74   \\
5075.8276  &  24375.46 & 30.57 &\vline &  5403.8058  &  24401.82  & 37.24   \\
5107.7807  &  24126.26 & 44.43 &\vline &  5404.8730  &  24261.93  & 41.39   \\
5108.8472  &  24079.34 & 39.29 &\vline &  5405.8598  &  24302.11  & 35.31   \\
5109.7623  &  24143.82 & 30.85 &\vline &  5407.8695  &  24314.65  & 34.22   \\
5109.7800  &  24221.74 & 27.24 &\vline &  5430.8126  &  24397.33  & 32.08   \\
5110.7724  &  24281.43 & 29.66 &\vline &  5432.7291  &  24217.71  & 40.97   \\
5112.7689  &  24325.34 & 28.45 &\vline &  5434.8370  &  24277.26  & 30.19   \\
5112.7808  &  24281.70 & 29.68 &\vline &  5436.8532  &  24391.30  & 29.11   \\
5113.7627  &  24226.34 & 32.90 &\vline &  5437.8192  &  24431.76  & 34.17   \\
5164.7436  &  24276.41 & 29.50 &\vline &  5491.5926  &  24425.56  & 46.26   \\
5166.7405  &  24373.37 & 34.20 &\vline &  5492.7290  &  24320.54  & 42.54   \\
5167.6315  &  24342.35 & 32.31 &\vline &  5493.7936  &  24351.56  & 32.04   \\
5168.6411  &  24288.16 & 21.91 &\vline &  5494.8416  &  24371.35  & 37.64   \\
5170.7375  &  24230.07 & 34.85 &\vline &  5497.8011  &  24298.56  & 30.49   \\
5172.6838  &  24297.89 & 41.50 &\vline &  5521.6567  &  24476.62  & 22.33   \\
5174.6691  &  24319.12 & 26.32 &\vline &  5523.6496  &  24471.39  & 25.01   \\
5181.5403  &  24428.12 & 32.69 &\vline &  5524.5791  &  24387.17  & 17.30   \\
5198.5285  &  24328.64 & 35.72 &\vline &  5525.6558  &  24390.15  & 17.25   \\
5199.6074  &  24340.87 & 28.77 &\vline &  5526.6492  &  24439.61  & 15.05   \\
5200.6219  &  24390.29 & 30.19 &\vline &  5527.6314  &  24449.32  & 25.68   \\
5201.5849  &  24400.51 & 21.81 &\vline &  5583.6321  &  24489.01  & 14.01   \\
5202.6165  &  24441.51 & 25.55 &\vline &  5584.5958  &  24630.11  & 21.68   \\
5203.5918  &  24288.39 & 30.66 &\vline &  5585.6161  &  24437.91  & 36.76   \\
5204.6389  &  24263.70 & 31.88 &\vline &  5587.6035  &  24290.41  & 50.00   \\
5254.5427  &  24299.59 & 41.61 &\vline &  5588.6275  &  24391.65  & 66.95   \\
5255.5272  &  24402.04 & 40.17 &\vline &             &            &	       \\

\hline
\hline

\end{array}
$$
\end{table*}
   
\subsection{Stellar rotation} 
\label{rotation}
A systematic investigation of the stellar activity is mandatory to avoid wrong 
interpretations of the observed RV variations. There are several possibilities to probe 
the stellar activity. The line profile asymmetry (bisector) and the \ion{Ca}{ii} lines are known 
as reliable stellar activity indicators.
These activity indicators, if they show periodic variations, 
can be used to determine the stellar rotation period. 
Besides the spectroscopic methods, photometric observations are also commonly used 
to find the stellar rotation period. 

\mbox{HIP 11952} itself is not a star with high stellar activity. 
This is inferred from the absence of emission cores in \ion{Ca}{ii} K ($\lambda$3934) 
and H ($\lambda$3967).
Furthermore, no H$\alpha$ emission line was observed in the spectra. 
The projected rotational velocity is also relatively 
low ($v \sin{i} = 5.2\, \mathrm{km\,s}^{-1}$) compared to other F-type 
dwarf stars (\cite{demedeiros99}). 
Finally, using the relation $P/\sin{i}= 2\pi\,R_{*}/v\sin{i}$, the maximum stellar rotation 
period is found to be 15.7$\pm$2.5 d.

\subsubsection{Line profile asymmetry}
\label{lineprofile}
\begin{figure}[h]
\centering
\includegraphics[width=8.5cm, height=8.5cm]{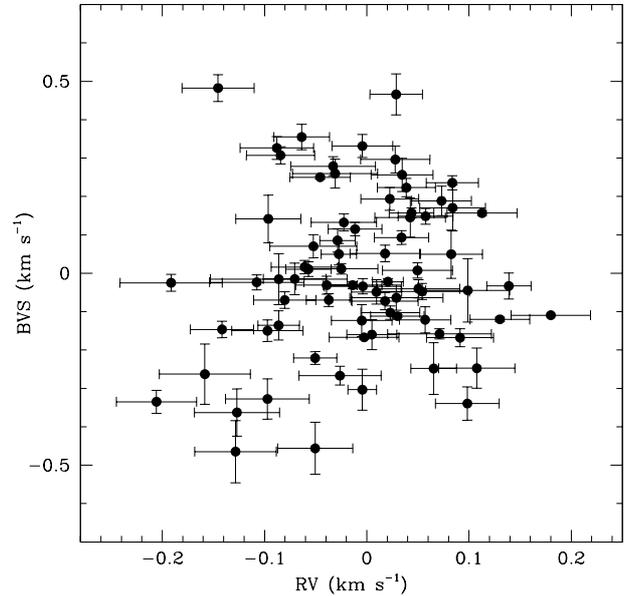}
\caption{The figure shows the measurements of BVS, ploted against residual RVs. The plot shows 
no correlation between BVS and RV. We computed a correlation coefficient $c=0.1$.}
\label{bvsrv}
\end{figure}

The line profile asymmetry can be quantified by the bisector velocity span (BVS). 
A definition of the BVS is given, e.g., in \cite{hatzes96}.  
We measured the BVS of the stellar spectra and searched for any periodicity that 
might be related to the RV variation.
In the GLS periodogram of BVS we found, however, no significant peak. 
Thus,  we cannot use the bisector to determine the stellar rotation period. 
We then searched for a correlation between BVS and RV to find out whether the observed RV variation 
is due to rotational modulation. We computed a correlation coefficient $c=0.1$ between the RV and BVS (Fig.~\ref{bvsrv}). 
This value indicates that the RV variation is not correlated with the BVS. 
However, it does not give any hint about the stellar rotation period.

\subsubsection{\ion{Ca}{ii} analysis}
\label{caII}
As mentioned before we found no emission cores in the \ion{Ca}{ii} H \& K lines 
($\lambda\lambda$3934,3968) which are, in general, excellent stellar activity 
indicators to probe the rotational modulation of the star. 
Nevertheless, we investigated to possibility to use \ion{Ca}{ii} K lines 
to find any indication of stellar activity. 
We calculated the stellar activity index, known as $S$-index, 
similar to the method described in \cite{vaughan78} and \cite{santos2000}.  
We found a significant periodicity of 36 d in the periodogram, but the 
error bar of the individual measurement is large. This value is also close 
to the typical observational window of about one month. Thus, also by considering 
$P/\sin{i}$ value, we do not adopt this as the stellar rotation period.

\begin{figure}[t]
\centering
\includegraphics[width=8.5cm, height=8.5cm]{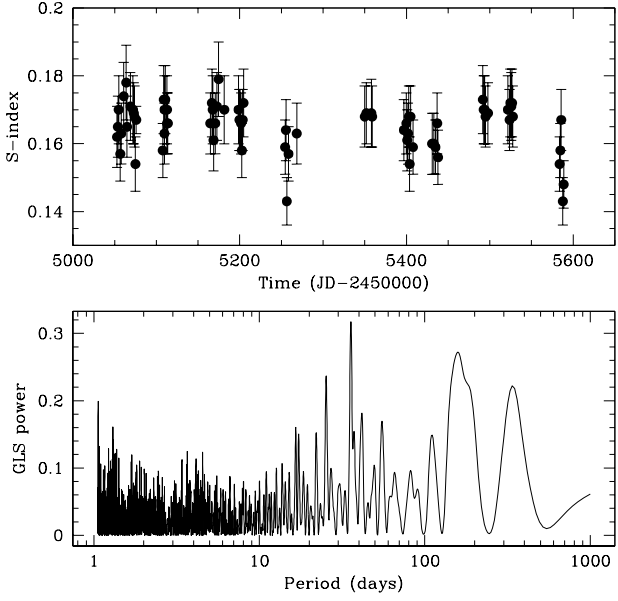}
\caption{Upper panel: The measurements of \ion{Ca}{ii} $\lambda$3934 $S$-index. 
Lower panel: The GLS periodogram of \ion{Ca}{ii} $\lambda$3934 shows 
a clear peak corresponding to a period of $P=36$ d. However, the error bar of 
the individual measurement is large and the period is close to the 
observational time window of about 1 month.}
\label{ca8662}
\end{figure}

We exploited the capability of FEROS to investigate 
the \ion{Ca}{ii} lines $\lambda\lambda$8498,8662, 
following the technique presented e.g. by \cite{larson93}, 
where \ion{Ca}{ii} $\lambda$8662 was used to determine the stellar rotation period.
\cite{setiawan10}, for example, used the equivalent width (EW) variations 
of \ion{Ca}{ii} $\lambda$8498 as a stellar activity indicator to estimate a stellar rotation period 
that agrees with the bisector analysis.

The EW variation of Ca $\lambda$8498 of \mbox{HIP 11952} indeed 
shows a periodic variation with $P=1.76$ d with a FAP of few percent. 
Thus, it is only marginaly significant. 
Following \cite{larson93} we then measured the EW variation of \ion{Ca}{ii} $\lambda$8662.
Interestingly, we found a significant periodicity in the EW variation 
of \ion{Ca}{ii} $\lambda$8662. The signal corresponds to a peak at a period of 
$P=4.82$ d with FAP$=8\times 10^{-3}$.
Fig.~\ref{ca8662} shows the EW variation and GLS periodogram of the \ion{Ca}{ii} $\lambda$8662 line. 
Assuming that this feature is related to the stellar magnetic activity caused by starspots, 
the period is most-likely linked to the stellar rotation.

\begin{figure}[t]
\centering
\includegraphics[width=8.5cm, height=8.5cm]{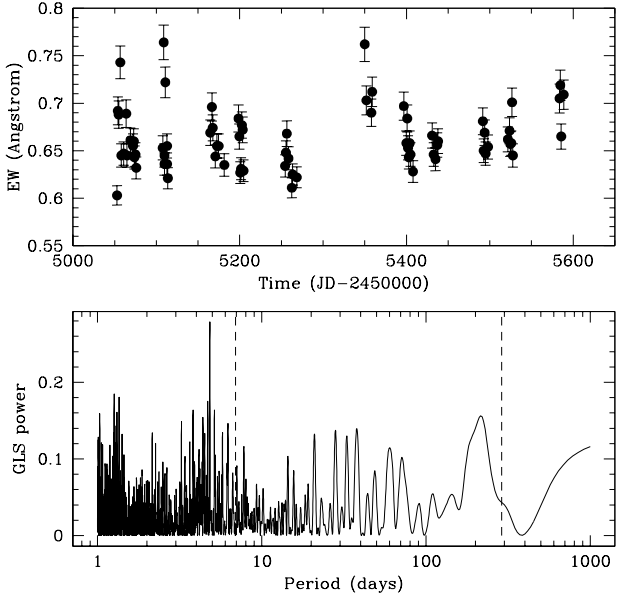}
\caption{Upper panel: The measurements of \ion{Ca}{ii} $\lambda$8662 EW variation. 
Lower panel: The GLS periodogram of \ion{Ca}{ii} $\lambda$8662 shows 
a clear peak corresponding to a period of $P=4.82$ d. The dashed line shows the 
period of the RV variations.}
\label{ca8662}
\end{figure}

\subsubsection{Photometric data}
\label{ph}
Photometric V band observations of \mbox{HIP 11952} are available in the Hipparcos catalogue.
Unfortunately, the data set is very sparse, with 72 photometric measurements over a time span of 893 d.
The minimum time-interval between data points is $\approx$\,0.014\,d, and the
typical photometric errors are $\approx$\,0.02 mag. Due to long-term gaps of several days in the data set, 
periodicities of a few days cannot be reliably detected.
The sampling allows for the detection of very short-term (few hours) as well as
long-term (20 d) variations.

We searched for periodicity in the photometric data using a combination of two periodogram analysis 
techniques: the Scargle periodogram (\cite{scargle82}) and the CLEAN
algorithm (\cite{roberts87}). The combination of these two techniques provides a reliable period
detection as outlined in several rotational period and variability studies
(e.g., \cite{rodriguezledesma09}).

\begin{figure}[h]
\centering
\includegraphics[width=8.5cm,height=5.5cm]{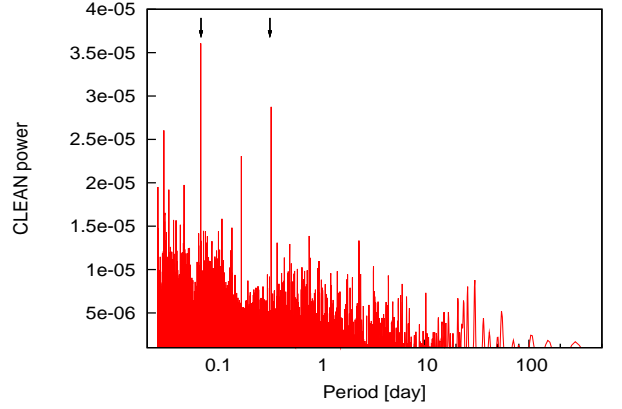}
\caption{The photometric observations of \mbox{HIP 11952} show two peaks at $P=0.3$ and $P=0.072$ d.}
\label{phot} 
\end{figure}

\begin{figure}[h]
\centering

\includegraphics[width=8.5cm,height=5.5cm]{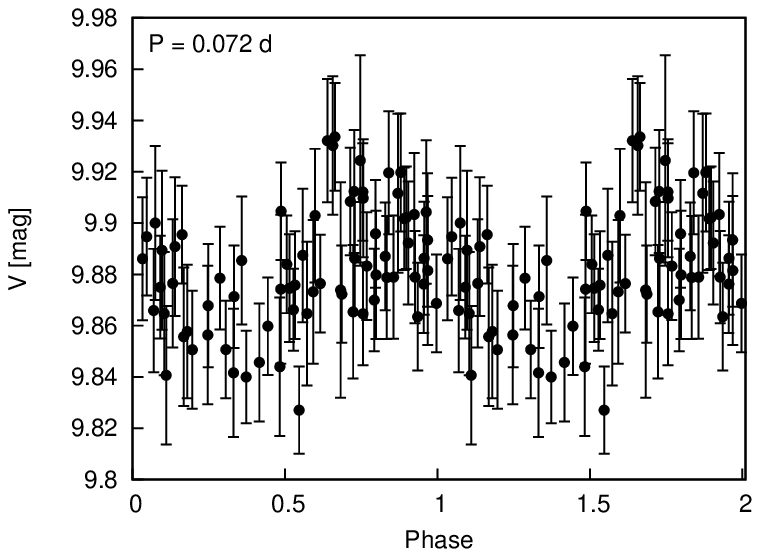}
\includegraphics[width=8.5cm,height=5.5cm]{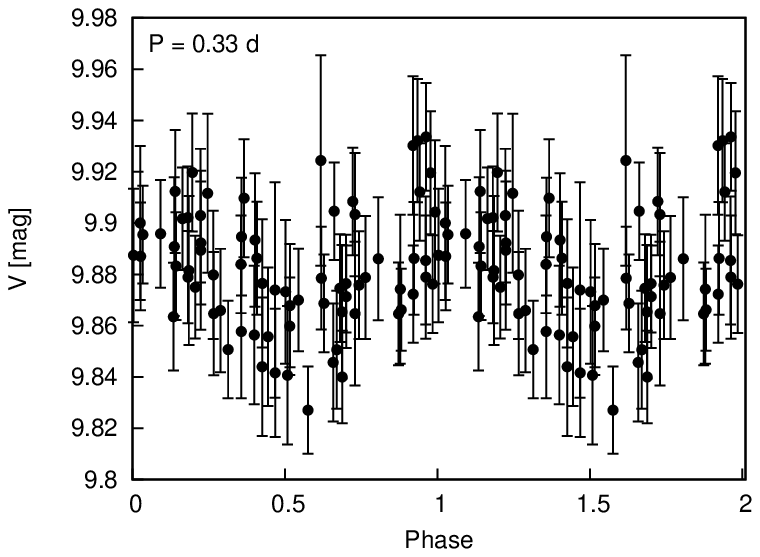}

\caption{Phase folded light curves from the Hipparcos photometric data.}
\label{plc} 
\end{figure}

Based on the Scargle periodogram, we detected significant signals at $P_1=$ 0.072, $P_2=$
0.33 and at $P_3=$ 2 d.  When the CLEAN algorithm is applied, the 2 d signal is removed and 
therefore, we concluded that it is probably a false peak or alias due to the clumpy data sampling. 
Both the 0.072 and 0.33 d periods in the power spectrum remains (Fig.~\ref{phot}), with FAP, 
based on the Scargle periodogram, of 0.5\,\% and 1\,\%, respectively. All
other peaks in Fig.\,6 have larger FAPs. We have also computed the statistical
F-test and a derived FAP from it (\cite{Scholz2004a} and
\cite{rodriguezledesma09}). The FAP$_{Ftest}$ represents the probability that the
period found is caused by variations in the photometric noise, and therefore it is
independent of the periodogram analysis. The FAP Ftest derived for the 0.072 and
0.33 d detected peaks are 5\,\% and 11\,\%, respectively. 
Fig.~\ref{plc} shows the phase folded light curves with the periods of 0.072 and 0.33 d.

Due to the quality of the data set, however, it is difficult to ensure the significance of
these periods. Our analysis suggests some evidence for short-term photometric
variations, which might be interpreted as possible pulsation modes in this F-type
star.

Based on the analysis of the stellar activity indicators, we assume 
a stellar rotation period of 4.8 d, as derived from the \ion{Ca}{ii} $\lambda$8662. 
Additionally, there might be evidence for stellar pulsations 
with intra-day periodicities
 
\section{Planetary companion}
\label{planet}
Since the long-period and short-period RV variations have different characteristics 
from those of the stellar activity indicators, we concluded that they are most likely caused by the 
presence of unseen companions.

\begin{figure}[h]
\centering
\includegraphics[width=8.5cm,height=7.0cm]{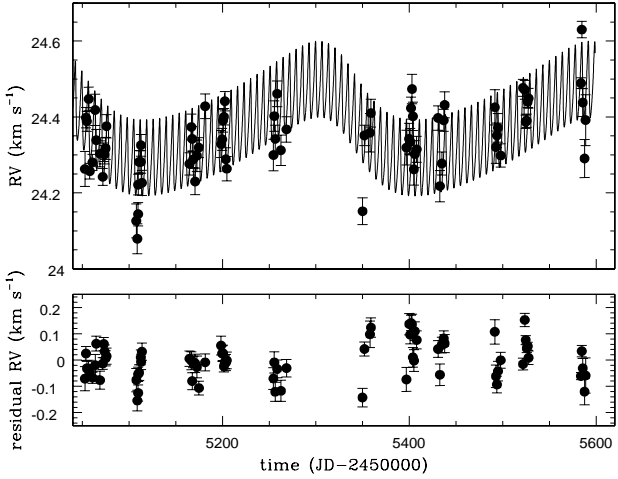}
\caption{Orbital solution of a two-planets Keplerian fit (upper panel). 
The orbital parameters are listed in Table.\ref{orbsol}. The residual 
velocities are shown in the lower panel.}
\label{orbkepler}
\end{figure}

We computed the orbital solution by using a two-components Keplerian model. 
In Fig.~\ref{orbkepler} we show the calculated orbital fit and residual velocities. 
The orbital paramaters are given in Table~\ref{orbsol}.
With a derived primary mass of 0.83 M$_\odot$, we calculated the minimum masses of 
the companions  $m_{2} \sin{i}=$ 0.78 M$_\mathrm{Jup}$ for the inner and
$m_{2} \sin{i}=$ 2.93 M$_\mathrm{Jup}$ for the outer companion.
The orbital semi-major axes are 0.07 AU and 0.81 AU, respectively.
The planetary orbits have moderate eccentricities of 0.35 and 0.27, which seem to be not unusual, 
based on the statistics of the eccentricity of exoplanets (see e.g., 
www.exoplanet.eu).

\begin{table}[h]
\caption{Orbital parameters for \mbox{HIP 11952} b and c} 
\label{orbsol}
\vspace{-0.5cm}
$$
\begin{array}{lcrrrr}
\hline
\hline
\mathrm{Parameter}& \mathrm{Unit}     & \mathrm{HIP\,11952 b}  & \mathrm{HIP\,11952 c}   \\
\hline
P		& \mathrm{d}          & 290.0\pm{16.2}         & 6.95\pm{0.01} 	 \\
T_{0}   	& \mathrm{JD}  	      & 5402.0\pm{1.3} 	       & 5029.2\pm{0.04}   \\
&-2450000       &                     & 		       &			 \\
e      		&		      & 0.27\pm{0.10}	       & 0.35\pm{0.24}  	 \\
\omega_{1}	& \mathrm{deg}        & 59.3\pm{2.5}	       & 61.2\pm{6.6}		 \\
K_{1}		& \mathrm{m\,s}^{-1}  & 105.2\pm{14.7}         & 100.3\pm{19.4}  	 \\
m_{2}\sin{i} 	& M_{\mathrm{Jup}}    & 2.93\pm{0.42}	       & 0.78\pm{0.16}  	 \\
a 		& \mathrm{AU}	      & 0.81\pm{0.02}	       & 0.07\pm{0.01}  	 \\
\hline

V_{0}		& \mathrm{km\,s}^{-1} & 24.365\pm{0.01}        &  	 \\ 
\sigma(O-C)	& \mathrm{m\,s}^{-1}  & 70.22		       &   	 \\
\mathrm{reduced\,} \chi^{2}&	      & 1.3  	       	       &         \\
\hline
\hline
\end{array}
$$
\end{table}

When calculating the orbital solutions, we obtained a relatively large $\sigma(O-C)$ value. 
A possible explanation to this is the presence of another unseen companion or RV jitter 
due to the interaction between the two companions.
Because of the absence of the emission cores in \ion{Ca}{ii} H \& K, the large $\sigma(O-C)$ 
is most likely not due to the intrinsic stellar variability.

We investigated the residual velocities and found a significant signal at $\sim$40 days. 
However, the amplitude of the residual RV variations is in the order of the error bars. 
Moreover, the window function shows also a peak close to this period.
To detect further low-amplitude RV variations, more intensive high-precise RV measurements 
are needed.

From the knowledge of $P_\mathrm{rot}/\sin{i}=15.7$ d (Table~\ref{par}) and the 
rotation period $P=4.82$ d, we derived an inclination angle of the 
stellar rotation of $18^\circ$. Assuming that the orbital inclination of the
companion does not differ much from the stellar rotation inclination angle, 
we estimated true companion masses of 2.5 and 9.5 $M_{\mathrm{Jup}}$.

\section{Discussion}
\label{discussion}
\begin{figure}[h]
\centering
\includegraphics[width=9cm]{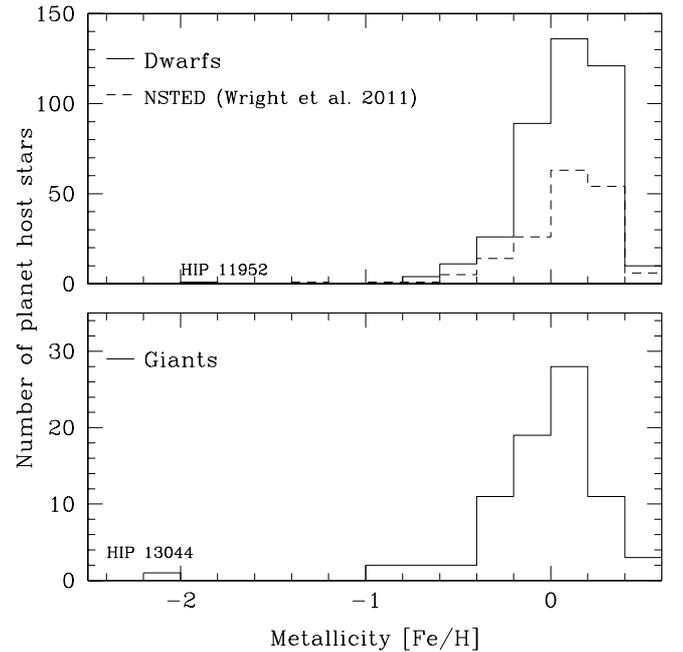}
\caption{The metallicity distribution of stars hosting planets. Solid lines in the top and 
bottom panels represent dwarfs and giants, respectively (data from 
the exoplanet encyclopedia 
www.exoplanet.eu). Dash lines show data from the   
exoplanet orbit database (\cite{wright11}) which does not discriminate between 
dwarf and giant stars.}
\label{metal}
\end{figure}

A fundamental parameter of \mbox{HIP 11952} is the stellar metallicity. 
The metallicity issue here is particularly interesting, since according to the theory of 
planet formation via core-accretion processes, planetary companions 
around such metal poor stars like \mbox{HIP 11952} are not expected.
The majority of the planet host main-sequence stars 
are metal rich (Fig.~ \ref{metal}, upper panel).

For dwarf stars, the metallicity generally reflects the 
initial metallicity at star formation. 
Assuming that \mbox{HIP 11952} is a dwarf or a turn off star, 
its initial low metallicity makes \mbox{HIP 11952} unusual among the planetary systems 
discovered so far (Fig.~\ref{metal}).

The both planetary companions around \mbox{HIP 11952} belong to only few planets that 
have been discovered in low metallicity systems ([Fe/H]$<$-1.5), comparable to \mbox{HIP 13044} 
reported by \cite{setiawan10}. 
Note that, so far also only few planets have been detected with 
host star's metallicities -1$<$[Fe/H]$<$-0.5. 
This group includes $\sim$60\% main-sequence and $\sim$40\% giants. 
From the current statistics, about 50\% of the giant planet-host stars 
have sub-solar metallicities. 

Whether the metallicity of a giant reflects 
its initial metallicity is still under debate. 
The convective envelopes in giant star are much more extended than in 
main-sequence stars (e.g. \cite{pasquini07}) and they might alter 
their surface chemical abundances.

The presence and formation of planets around metal-poor stars, in particular those 
with metallicities [Fe/H]$<-1.5$ are still poorly understood. 
According to the core accretion theory (e.g., \cite{safronov69}; \cite{pollack96}) 
high metallicity is required for the formation of planets. 
Alternatively, planets around metal-poor stars could form by gravitational disk instability processes 
(e.g., \cite{boss98}) or other mechanisms (see e.g., \cite{nayakshin11} 2011). 
According to \cite{nayakshin11} (2011) planets around such metal-poor stars have 
no solid cores and may form as a result of the second collapse 
(H2 disassociation) of their embryos and radial migrations. 
Further discoveries of planets around metal-poor stars can provide 
more constraints on currently different planet formation theories.

The planets around \mbox{HIP 11952} have orbits with moderate eccentricities.
To examine the dynamical stability of the system, numerical simulations are necessary to 
find stable planetary configurations, see e.g., \cite{barnes2004}.
Following their calculations, systems with two or more planets in large separated 
orbits are fully stable.
The orbits of \mbox{HIP 11952} b and c are in a 42:1 ratio and thus far beyond the 
10:1 resonance. Therefore, the system \mbox{HIP 11952} is most likely fully stable.

Finally, \mbox{HIP 11952} and its planets are among the oldest planetary systems known. 
\mbox{HIP 11952} is also older than typical stellar ages in the Galactic 
thick disk. A possible connection to a metal-poor stellar
stream reported by \cite{arifyanto06} is an interesting aspect 
since it suggests that \mbox{HIP 11952} might actually belong to a part
of the thick disk that has been accreted from a disrupted former satellite galaxy, similar to \mbox{HIP 13044}.
The age estimation of $12.8\pm2.6$ Gyr given by \cite{feltzing2001} is 1 Gyr older 
than that of \mbox{HD 114752} which has an estimated age of $11.8\pm3.9$ Gyr.
\mbox{HIP 11952}'s age is close to the one of
\mbox{HE 1523-0901}, which is the oldest star ($13.2\pm2.7$ Gyr) 
known so far (\cite{frebel07}). 
The old stellar age is supported by the very low metallicity of \mbox{HIP 11952} 
which is typical of Population II stars.


\section{Conclusions}
\label{con}
We observed RV variations of \mbox{HIP 11952}. The spectroscopic and 
photometric analysis of the star show that the periodic RV variations are not caused by 
the intrinsic stellar variability.
Based on our analysis, we detected two planetary companions  
around the metal poor star \mbox{HIP 11952} with orbital 
periods of 6.95 d and 290 d. 
We found evidence for intra-day stellar pulsations and observed a 
stellar rotation of 4.82 d. 
We computed the companion's minimum mass of $m_{2} \sin{i}=0.78$ M$_\mathrm{Jup}$ for 
the inner planet and $m_{2} \sin{i}=2.93$ M$_\mathrm{Jup}$ for the outer one. 
Additional high-precise RV measurements are necessary to improve the orbital 
solution and put more constraints on the eccentricities. 
Further RV observations might also reveal the presence 
of other low-mass companions in the system. 
From the metal abundance analysis that we
carried out, we obtained an average [Fe/H]$=-1.90\pm0.06$ 
from \ion{Fe}{i} and \ion{Fe}{ii}, respectively, 
which makes \mbox{HIP 11952} b and c the first 
planets discovered around a dwarf or subgiant star with [Fe/H]$<$-1.5. 
This discovery is also remarkable since the planetary system most likely 
belong to the first generation of planetary systems in the Milky Way.

\begin{acknowledgements}
We thank N. Christlieb for the support and useful discussions. 
We also thank A. Mueller, M. Zechmeister, J. Carson, W. Wang, C. Ruhland, 
T.~S. Hartung, S. Albrecht, R. Lachaume and T. Anguita 
for observing HIP 11952 with FEROS. 
\end{acknowledgements}

\end{document}